\documentclass{PoS}
\usepackage{setspace}
\usepackage{slashed}
\usepackage{verbatim}    
\usepackage{graphicx} 
\usepackage{wrapfig}
\usepackage{dsfont}
\usepackage{epsfig}
\usepackage{epstopdf}
\usepackage{amsmath}
\usepackage{amsfonts,amssymb}
\usepackage{cleveref}
\usepackage{xcolor}

\newcommand{\be}{\begin{equation}}
\newcommand{\ee}{\end{equation}}
\newcommand{\bea}{\begin{eqnarray}} 
\newcommand{\eea}{\end{eqnarray}}
\newcommand{\MSbar}{{\overline{\rm MS}}}

\newcommand{\la}{\lambda}

\newcommand{\gtilde}{\frac{g^2}{16 \, \pi^2}\; }
\title{Fine-Tuning of the Yukawa and Quartic Couplings in Supersymmetric QCD}
\ShortTitle{Yukawa and Quartic Couplings in Supersymmetric QCD}

    \author{M.~Costa$^{a,\, b}$, \speaker{H.~Herodotou}$^{ \,a}$, H.~Panagopoulos$^{\,a}$\\
	\llap{}$^a$Department of Physics, University of Cyprus, Nicosia, CY-1678, Cyprus\\
	$^b$Department of Chemical Engineering, Cyprus University of Technology, \\ 30 Archbishop Kyprianou Str., 3036, Limassol, Cyprus \\
	{\rm E-mail}:  \email{kosta.marios@ucy.ac.cy}, \email{herodotos.herodotou@ucy.ac.cy}, \email{panagopoulos.haris@ucy.ac.cy}}
	
\abstract{In this work, we investigate the fine tuning of parameters in $\mathcal{N} = 1$ Supersymmetric QCD, discretized on a Euclidean lattice. Specifically, we study the renormalization of the Yukawa (gluino-quark-squark interactions) and the quartic (four-squark interactions) couplings. At the quantum level, these interactions suffer from mixing with other operators which have the same transformation properties. We exploit the symmetries of the action, such as charge conjugation and parity, in order to reduce the allowed mixing patterns. To deduce the renormalizations and the mixing coefficients we compute, perturbatively to one-loop and to the lowest order in the lattice spacing, the relevant three-point and four-point Green’s functions using both dimensional and lattice regularizations. Our lattice formulation involves the Wilson discretization for the gluino and quark fields; for gluons we employ the Wilson gauge action; for scalar fields (squarks) we use na\"ive discretization. We obtain analytic expressions for the renormalization and mixing coefficients of the Yukawa couplings; they are functions of the number of colors $N_c$, the gauge parameter $\alpha$, and the gauge coupling $g$. Furthermore, preliminary results on the quartic couplings are also presented.
\begin{center}
\includegraphics[scale=0.45]{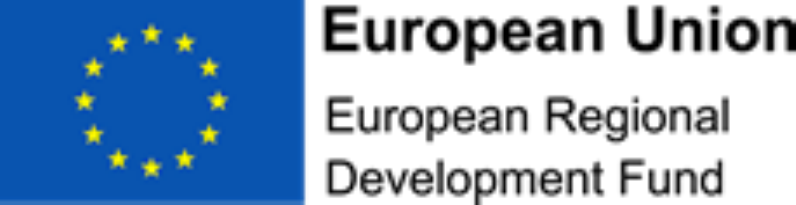}
\includegraphics[scale=0.45]{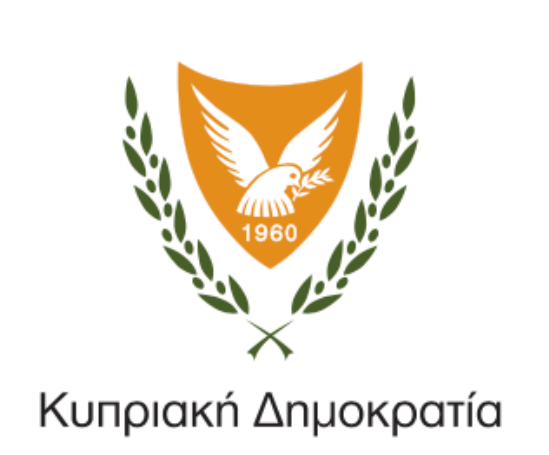}
\includegraphics[scale=0.45]{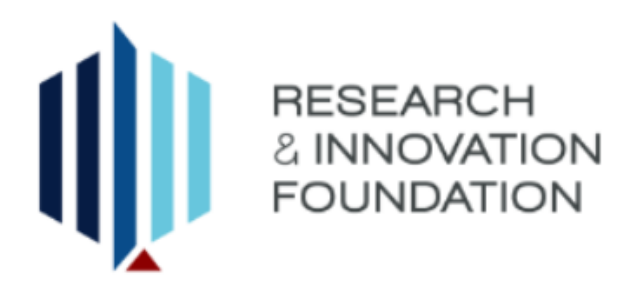}
\includegraphics[scale=0.45]{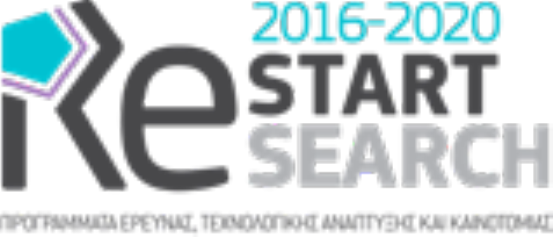}
\end{center}}

\FullConference{%
The 39th International Symposium on Lattice Field Theory,\\
8th-13th August, 2022,\\
Rheinische Friedrich-Wilhelms-Universität Bonn, Bonn, Germany
}

\begin{document}

	\maketitle
	
	\section{Introduction}

Supersymmetric models of strongly coupled theories are a very promising candidate for new physics beyond the standard model. In  recent  years,  numerical  lattice  studies  of  supersymmetric extensions  of QCD are becoming  within  reach. However, there are several well-known obstacles arising from the breaking of Supersymmetry in a regularized theory on the lattice, including the necessity for fine tuning of the theory's bare Lagrangian. We are addressing these problems via perturbative calculations, to one loop and to lowest order in the lattice spacing. The quantities, which we calculate in this work, are important ingredients in extracting nonperturbative information for Supersymmetric Theories through lattice simulations. 

As is the case with all known regulators, lattice breaks supersymmetry explicitly but it is the only regulator which describes many aspects of strong interactions nonperturbatively. Note that the coupling constants appearing in the lattice action are not all identical. On one hand, gauge invariance of the action dictates that some of the interaction parts will have the same coupling, $g$ (gauge coupling); this is the case for the kinematic terms which contain covariant derivatives and thus gluons couple with quarks, squarks, gluinos and other gluons with the same gauge coupling constant. The Yukawa interactions between quarks, squarks and gluinos as well as the four-squark interactions contain a different coupling constant, which must be fine tuned on the lattice. Exploiting the symmetries of the action, we reduce the number of possible interaction terms and therefore, their tuning. It is desirable to employ a lattice discretization which preserves as many as possible of the continuum symmetries, so that the relevant operators to be tuned will be fewer. The overlap formulation can be used, in order to preserve chiral symmetry, but we will first investigate these tunings using the Wilson fermion action. The use of the overlap action is beyond the scope of the present work.

In this work, we present a number of different results, which are obtained by using the SQCD action on the lattice, regarding the renormalization of the Yukawa and quartic couplings. For the gluon fields we use the Wilson gauge action, for fermions (quark, gluino fields) we employ the Wilson fermion action, and for the squark fields we use na\"ive discretization. After presenting the basics of the computation setup (Section \ref{comsetUP}), we start with a discussion of the renormalization of the Yukawa couplings (Section \ref{couplingY}) both in dimensional and lattice regularizations. We utilize the $\MSbar$ renormalization scheme and we determine the renormalization factors to one-loop order. In the same way, we also present some preliminary results for the renormalization of the quartic couplings (Section \ref{couplingQ}). Finally, we end with a short summary and our future plans (Section~\ref{summary}).
	
	\section{Formulation and Computational Setup}
	\label{comsetUP}
In our calculation we make use of the SQCD action in the Wess-Zumino (WZ) gauge \cite{Curci:1987, Schaich:2014, Giedt:2014, Bergner:2016, Costa:2017rht, Costa:2018mvb}. In our lattice calculation, quarks ($\psi$), squarks ($A_\pm$) and gluinos ($\lambda$) are defined on the lattice points whereas gluons ($u_\mu$) are defined on the links between adjacent points: $U_\mu (x) = \text{exp}[i g a T^{\alpha} u_\mu^\alpha (x+a\hat{\mu}/2)]$; $\alpha$ is a color index in the adjoint representation of the gauge group. For Wilson-type quarks and gluinos, the Euclidean action ${\cal S}^{L}_{\rm SQCD}$ on the lattice and in the massless limit becomes:    
\begin{small}
\bea
{\cal S}^{L}_{\rm SQCD} & = & a^4 \sum_x \Big[ \frac{N_c}{g^2} \sum_{\mu,\,\nu}\left(1-\frac{1}{N_c}\, {\rm Tr} U_{\mu\,\nu} \right ) + \sum_{\mu} {\rm Tr} \left(\bar \lambda  \gamma_\mu {\cal{D}}_\mu\lambda  \right ) - a \frac{r}{2} {\rm Tr}\left(\bar \lambda   {\cal{D}}^2 \lambda  \right) \nonumber \\ 
&+&\sum_{\mu}\left( {\cal{D}}_\mu A_+^{\dagger}{\cal{D}}_\mu A_+ + {\cal{D}}_\mu A_- {\cal{D}}_\mu A_-^{\dagger}+ \bar \psi  \gamma_\mu {\cal{D}}_\mu \psi  \right) - a \frac{r}{2} \bar \psi   {\cal{D}}^2 \psi  \nonumber \\
&+&i \sqrt2 g_Y \big( A^{\dagger}_+ \bar{\lambda}^{\alpha}  T^{\alpha} P_+ \psi   -  \bar{\psi}  P_- \lambda^{\alpha}   T^{\alpha} A_+ +  A_- \bar{\lambda}^{\alpha}  T^{\alpha} P_- \psi   -  \bar{\psi}  P_+ \lambda^{\alpha}   T^{\alpha} A_-^{\dagger}\big)\nonumber\\  
&+& \frac{1}{2} g^2_4 (A^{\dagger}_+ T^{\alpha} A_+ -  A_- T^{\alpha} A^{\dagger}_-)^2 
\Big],
\label{susylagrLattice}
\eea
\end{small}where: $P_\pm= (1 \pm \,\gamma_5)/2$, $U_{\mu \nu}(x) =U_\mu(x)U_\nu(x+a\hat\mu)U^\dagger_\mu(x+a\hat\nu)U_\nu^\dagger(x)$, $a$ is the lattice spacing, and a summation over flavors is understood in the last three lines of Eq.~(\ref{susylagrLattice}). The 4-vector $x$ is restricted to the values $x = na$, with $n$ being an integer 4-vector. The definitions of the covariant derivatives are the standard definitions as shown in \cite{Costa:2017rht}. The terms proportional to the Wilson parameter, $r$, eliminate the problem of fermion doubling, at the expense of breaking chiral invariance. In the limit $a \to 0$ the lattice action reproduces the continuum one. In order to recover SUSY in the classical continuum limit, the tree level values of $g_Y$ and $g_4$ must coincide with $g$. 

Note that a discrete version of a gauge-fixing term, together with the compensating ghost field term, must be added to the action, in order to avoid divergences from the integration over gauge orbits; these terms are the same as in the non-supersymmetric case. Similarly, a standard “measure” term must be added to the action, in order to account for the Jacobian in the change of integration variables: $U_\mu \to u_\mu$\,. 

In Refs.~\cite{Costa:2017rht} and~\cite{Costa:2018mvb}, first lattice perturbative computations in the context of SQCD were presented; apart from the Yukawa and quartic couplings, the renormalization of all parameters and fields appearing in the supersymmetric Lagrangian using Wilson gluons and fermions have been extracted. In addition, the mixing of some composite operators under renormalization has been explored. The results in these references~\cite{Costa:2017rht,Costa:2018mvb} will find further use in the present work. The additional calculations of the fine tunings for the gluino-quark-squark and four-squark couplings are essential prerequisites towards nonperturbative investigations. 

Parity ($\cal{P}$) and charge conjugation  (${\cal{C}}$) are symmetries of the lattice action and their definitions are presented below:
\be
{\cal{P}}:\left \{\begin{array}{ll}
&\hspace{-.3cm} U_0(x)\rightarrow U_0(x_{P})\, ,\qquad U_k(x)\rightarrow U_k^{\dagger}(x_{P}-a\hat{k})\, ,\qquad k=1,2,3\\
&\hspace{-.3cm} \psi_f(x)\rightarrow \gamma_0  \psi_f(x_{P})\\
&\hspace{-.3cm}\bar{ \psi}_f(x) \rightarrow\bar{ \psi}_f(x_{P})\gamma_0\\
&\hspace{-.3cm} \la_f(x)\rightarrow \gamma_0  \la_f(x_{P})\\
&\hspace{-.3cm}\bar{ \la}_f(x) \rightarrow\bar{ \la}_f(x_{P})\gamma_0\\
&\hspace{-.3cm} A_\pm(x) \rightarrow A_\mp^\dagger(x_{P})\\
&\hspace{-.3cm} A_\pm^\dagger(x) \rightarrow A_\mp(x_{P})
\end{array}\right .
\label{Parity}
\ee
where $x_{P}=(-{\bf{x}},x_0)$.

\be
{\mathcal {C}}:\left \{\begin{array}{ll}
&\hspace{-.3cm}U_\mu(x)\rightarrow U_\mu^{\star}(x)\, ,\quad \mu=0,1,2,3\\
&\hspace{-.3cm}\psi(x)\rightarrow -C \bar{ \psi}(x)^{T}\\
&\hspace{-.3cm}\bar{\psi}(x)\rightarrow{\psi}(x)^{T}C^{\dagger}\\
&\hspace{-.3cm} \la(x) \rightarrow C \bar{\la}(x)^{T}\\
&\hspace{-.3cm}\bar{\la}(x) \rightarrow -{\la}(x)^{T}C^{\dagger}\\
&\hspace{-.3cm}A_\pm(x) \rightarrow A_\mp(x) \\
&\hspace{-.3cm}A_\pm^\dagger(x) \rightarrow A_\mp^\dagger(x)
\end{array}\right . 
\label{Chargeconjugation}
\ee
where $^{\,T}$ means transpose. The matrix $C$ satisfies: $(C \gamma_{\mu})^{T}= C \gamma_{\mu}$, $C^T=-C$ and $C^{\dagger} C=1$. In 4 dimensions, in a standard basis for $\gamma$ matrices, in which $\gamma_0,\ \gamma_2$ ($\gamma_1,\ \gamma_3$) are (anti-)symmetric, $C = - {\rm i} \gamma_0 \gamma_2$. 

Further symmetries of the classical action are: \\
$U(1)_R$ rotates the quark and gluino fields in opposite direction:
\be
{\cal{R}}:\left \{\begin{array}{ll}
&\hspace{-.3cm} \psi_f(x)\rightarrow e^{i \theta \gamma_5}  \psi_f(x)\\
&\hspace{-.3cm}\bar{ \psi}_f(x) \rightarrow\bar{ \psi}_f(x)e^{i \theta \gamma_5}\\
&\hspace{-.3cm} \la(x)\rightarrow e^{-i \theta \gamma_5}  \la(x)\\
&\hspace{-.3cm}\bar{ \la}(x) \rightarrow\bar{ \la}(x)e^{-i \theta \gamma_5}
\end{array}\right.
\label{Rsym}
\ee \\
$U(1)_A$ rotates the squark and the quark fields in the same direction as follows: 
\be
{\cal{\chi}}:\left \{\begin{array}{ll}
&\hspace{-.3cm} \psi_f(x)\rightarrow e^{i \theta \gamma_5}  \psi_f(x)\\
&\hspace{-.3cm}\bar{ \psi}_f(x) \rightarrow\bar{ \psi}_f(x)e^{i \theta \gamma_5}\\
&\hspace{-.3cm} A_\pm(x) \rightarrow e^{ i \theta} A_\pm(x)\\
&\hspace{-.3cm} A_\pm^\dagger(x) \rightarrow e^{- i \theta} A_\pm^\dagger(x)
\end{array}\right .
\label{chiral}
\ee

For the purpose of studying Yukawa couplings, we examine the behavior under $\cal{P}$ and $\cal{C}$ of all gauge invariant dimension-4 operators having one gluino, one quark and one squark field, as shown in Table~\ref{tb:Ycoupling}. Note that all operators that we consider here are flavor singlets.

\begin{table}[ht]
\begin{center}
\scalebox{0.8}{
  \begin{tabular}{ c | c | c}
\hline \hline
Operators & $\cal{C}$ &$\cal{P}$ \\ [0.5ex] \hline\hline
    $\,A^{\dagger}_+ \bar{\lambda} P_+ \psi \, $&$-\bar{\psi}  P_+ \lambda A_-^{\dagger}$ & $\, A_- \bar{\lambda} P_- \psi \,$  \\[0.75ex]
    \hline
    $\,\bar{\psi}  P_- \lambda A_+ $&$-A_- \bar{\lambda} P_- \psi \,$ & $ \,\bar{\psi}  P_+ \lambda A_-^{\dagger} \,$  \\[0.75ex]
    \hline
    $\,A_- \bar{\lambda} P_- \psi  $&$-\bar{\psi}  P_- \lambda A_+ \,$ & $\, A^{\dagger}_+ \bar{\lambda} P_+ \psi \, $  \\[0.75ex]
    \hline 
    $\, \bar{\psi}  P_+ \lambda A_-^{\dagger}$&$-A^{\dagger}_+ \bar{\lambda} P_+ \psi \,$ & $ \,\bar{\psi}  P_- \lambda A_+ \, $  \\[0.75ex]
    \hline
$\,A^{\dagger}_+ \bar{\lambda} P_- \psi \, $&$-\bar{\psi}  P_- \lambda A_-^{\dagger}$ & $\, A_- \bar{\lambda} P_+ \psi \,$  \\[0.75ex]
    \hline
    $\,\bar{\psi}  P_+ \lambda A_+ $&$-A_- \bar{\lambda} P_+ \psi \,$ & $ \,\bar{\psi}  P_- \lambda A_-^{\dagger} \,$  \\[0.75ex]
    \hline
    $\,A_- \bar{\lambda} P_+ \psi  $&$-\bar{\psi}  P_+ \lambda A_+ \,$ & $\, A^{\dagger}_+ \bar{\lambda} P_- \psi \, $  \\[0.75ex]
    \hline 
    $\, \bar{\psi}  P_- \lambda A_-^{\dagger}$&$-A^{\dagger}_+ \bar{\lambda} P_- \psi \,$ & $ \,\bar{\psi}  P_+ \lambda A_+ \, $  \\[0.75ex]
    \hline

\hline
\end{tabular}}
\caption{Gluino-quark-squark operators which are gauge invariant, flavor singlet and with dimensionality 4. All matter fields carry a flavor index.}
\label{tb:Ycoupling}
\end{center}
\end{table}

There are two linear combinations of Yukawa-type operators which are invariant under $\cal{P}$ and $\cal{C}$ \cite{Giedt:2009}:
\bea
\label{chiInv}
&&A^{\dagger}_+ \bar{\lambda} P_+ \psi   -  \bar{\psi}  P_- \lambda A_+ +  A_- \bar{\lambda} P_- \psi   -  \bar{\psi}  P_+ \lambda A_-^{\dagger} \\
&&A^{\dagger}_+ \bar{\lambda} P_- \psi   -  \bar{\psi}  P_+ \lambda A_+ +  A_- \bar{\lambda} P_+ \psi   -  \bar{\psi}  P_- \lambda A_-^{\dagger}
\label{chiNonInv}
\eea
Thus, all terms within each of the combinations in Eqs.~(\ref{chiInv}) and~(\ref{chiNonInv}) are multiplied by the same Yukawa coupling, $g_{Y_1}$ and $g_{Y_2}$, respectively. 

In the case of a theory with massive quarks, ${\cal{R}}$ is not a symmetry. In the absence of anomalies, ${\cal \chi} \times {\cal R}$ leaves invariant each of the four constituents of the Yukawa term (Eq.~(\ref{chiInv})), but it changes the constituents of the ``mirror'' Yukawa term (i.e. a term with all $P_+$ and $P_-$ interchanged) by phases $e^{2i \theta}$ and $e^{-2i \theta}$ and thus, it guarantees the absence of a ``mirror'' Yukawa term.

	\section{Renormalization of the Yukawa coupling}
	\label{couplingY}
\begin{figure}[ht!]
\centering
\includegraphics[scale=0.28]{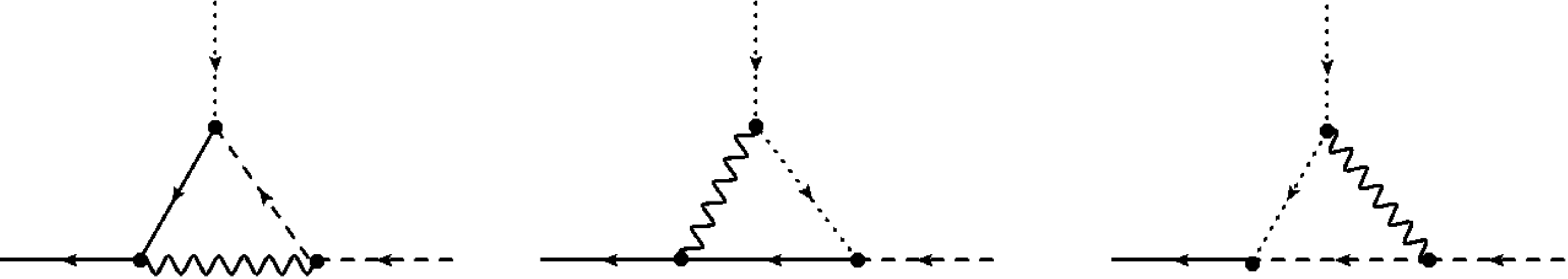}
\caption{One-loop Feynman diagrams leading to the fine tuning of $g_Y$. A wavy (solid) line represents gluons (quarks). A dotted (dashed) line corresponds to squarks (gluinos). 
In the above diagrams the directions of the external line depend on the particular Green's function under study. An arrow entering (exiting) a vertex denotes a $\la, \psi, A_+, A_-^{\dagger}$ ($\bar \la, \bar \psi, A_+^{\dagger}, A_-$) field. Squark lines could be further marked with a $+$($-$) sign, to denote an $A_+ \, (A_-)$ field.  
}
\label{couplingYukawa}
\end{figure}

There are three one-loop Feynman diagrams that enter the computation of the 3-point amputated Green's functions for the Yukawa couplings, shown in Fig. \ref{couplingYukawa}. We compute, perturbatively, the relevant three-point (3-pt) Green's functions using both dimensional regularization ($DR$) in $D = 4 - 2\epsilon$ dimensions and lattice regularization ($LR$).  

For the renormalization of $g_Y$, we impose renormalization conditions which result in the cancellation of divergences in the corresponding bare 3-pt amputated Green's functions with external gluino-quark-squark fields and thus, the renormalization factors are defined in such a way as to remove all divergences. The application of the renormalization factors on the bare Green's functions leads to the renormalized Green's functions, which are independent of the regulator ($\epsilon$ in $DR$, $a$ in $LR$). 

For convenience of computation, we are free to make appropriate choices of the external momenta. Having checked that no superficial IR divergences will be generated, we calculate the corresponding diagrams by setting to zero only one of the external momenta.
We present the one-loop Green's function for the Yukawa coupling for zero gluino momentum in $DR$ with external squark field $A_+$.
\begin{small}
\begin{align}
\label{gfA+}
\langle   &\lambda^{\alpha_1}(0)  {\bar{\psi}} (q_2) A_+(q_3)  \rangle^{DR, {\rm{1loop}}} = -i\,(2\pi)^4 \delta(q_2-q_3) \frac{g_Y g^2}{16\pi^2} \frac{1}{4\sqrt2 N_c} T^{\alpha_1} \times \nonumber\\
\hspace{-3.75cm}  &\Bigg[-3 (1 + \gamma_5) + ((1 + \alpha)(1 + \gamma_5) + 8 \gamma_5 c_{\rm hv}) N_c^2  + (1 + \gamma_5) (-\alpha + (3 + 2 \alpha) N_c^2 ) \left(\frac{1}{\epsilon}+ \log\left(\frac{\bar \mu^2}{q_2^2}\right) \right)\Bigg]
\end{align}
\end{small}where  $c_{\rm hv} = 0, 1$ for the na\"ive and 't Hooft-Veltman (HV) prescription \cite{Costa:2017rht} of $\gamma_5$. Note that the pole parts do not depend on $c_{\rm hv}$. We have verified that different choices for the external momenta lead to the same values for the coefficients of the pole parts. The Green's functions which involve the external squark fields $A_+^{\dagger}$, $A_-$ and $A_-^{\dagger}$ are similar to Eq. (\ref{gfA+}). The need of further conversion factors, which connect $\MSbar$ renormalized Green's functions to SUSY invariant Green's functions, is indicated by the supersymmetric Ward Identities~\cite{Wellegehausen:2018opt, Bergner:2018, Montvay:2002}. The calculation of these conversion factors is a purely continuum calculation; the same conversion factors can be applied to the renormalization functions extracted in $LR$.

The difference between the renormalized Green's functions and the corresponding Green's functions regularized on the lattice allows us to deduce the one-loop lattice renormalizations factors. The definition of the renormalization factors of the fields and the gauge coupling constant are the following:
\bea
\psi \equiv \psi^B &=& Z_\psi^{-1/2}\, \psi^R,\\
u_{\mu} \equiv u_{\mu}^B &=& Z_u^{-1/2}\,u^R_{\mu},\\
\la \equiv \la^B &=& Z_\la^{-1/2}\,\la^R,\\
c \equiv c^B &=& Z_c^{-1/2}\,c^R, \\
g \equiv g^B &=& Z_g^{-1}\,\mu^{\epsilon}\,g^R \label{g}, 
\eea
where $B$ stands for the bare and $R$ for renormalized quantities and $\mu$ is an arbitrary scale with dimensions of inverse length. For one-loop calculations, the distinction between $g^R$ and $g^B$ is inessential in many cases; we will simply use $g$ in those cases. 
The Yukawa coupling is renormalized as follows: 
\be
\label{gy}
g_Y \equiv g_Y^B = Z_Y^{-1} Z_g^{-1} \mu^\epsilon g^R,
\ee
where at the lowest perturbative order $Z_g Z_Y = 1$, and the renormalized Yukawa coupling coincides with the gauge coupling. \\
The components of the squark fields may mix at the quantum level, via a $2\times2$ mixing matrix ($Z_A$). We define the renormalization mixing matrix for the squark fields as follows:
\be
\label{condS}
\left( {\begin{array}{c} A^R_+ \\ A^{R\,\dagger}_- \end{array} } \right)= \left(Z_A^{1/2}\right)\left( {\begin{array}{c} A^B_+ \\ A^{B\,\dagger}_- \end{array} } \right).
\ee
In Ref.~\cite{Costa:2017rht} we found that in the $DR$ and $\MSbar$ scheme this $2\times2$ mixing matrix is diagonal. On the lattice, however, this matrix is non diagonal and the component $A_+ (A_-)$ mixes with $A_-^\dagger (A_+^\dagger)$. \\
Taking as an example the Green's function in $DR$ with external squark field $A_+$, the renormalization condition up to $g^2$ will be given by:
\be
\langle   \lambda(q_1)  {\bar{\psi}} (q_2) A_+(q_3) \rangle \Big \vert^\MSbar = Z_\psi^{-1/2} Z_\la^{-1/2}  (Z_A^{-1/2})_{++} \langle   \lambda(q_1)  {\bar{\psi}} (q_2) A_+(q_3) \rangle \Big \vert^{\rm{bare}} 
\label{renormC}
\ee
All appearances of coupling constants in the right-hand side of Eq.~(\ref{renormC}) must be expressed in terms of their renormalized values, via Eqs.~(\ref{g}-\ref{gy}). The left-hand side of Eq.~(\ref{renormC}) is just the $\MSbar$ (free of pole parts) renormalized Green's function. Similar to Eq.~(\ref{renormC}), the other renormalization conditions which involve the external squark fields $A_+^\dagger, A_-, A_-^\dagger$ are understood. We recall the following fundamental results on the renormalization factor in $DR$ which appear in the right-hand side of Eq.~(\ref{renormC}).
\bea
Z_\psi^{DR,\MSbar} &=& 1 + \frac{g^2\,C_F}{16\,\pi^2} \frac{1}{\epsilon}\left( 2 + \alpha \right), \quad \quad \quad
Z_{A}^{DR,\MSbar} = \bigg (1 + \frac{g^2\,C_F}{16\,\pi^2} \frac{1}{\epsilon}\left(1 + \alpha \right) \bigg) \mathds{1}\\
Z_{\lambda}^{DR,\MSbar} &=&  1 + \frac{g^2\,}{16\,\pi^2} \frac{1}{\epsilon} \left(\alpha\, N_c + N_f \right), \quad 
Z_{g}^{DR,\MSbar}=1 + \frac{g^2\,}{16\,\pi^2} \frac{1}{\epsilon} \left(\frac{3}{2} N_c - \frac{1}{2}N_f \right)
\eea
By means of Eq.~(\ref{renormC}) and for all Green's functions and all choices of the external momenta which we consider, we obtain the same value of $Z_Y^{DR, \MSbar}$:
\be
Z_Y^{DR, \MSbar}= 1 + \frac{g^2 N_c}{16\pi^2}\, \frac{1}{\epsilon}\, \frac{3}{2}\, C_F
\ee
where $C_F=(N_c^2-1)/(2\,N_c)$ is the quadratic Casimir operator in the fundamental representation. As expected from general renormalization theorems, the $\MSbar$ renormalization factors for gauge invariant objects are gauge-independent, as in the case of $Z_Y^{DR, \MSbar}$. 

On the lattice, the renormalization matrix $Z_A$ is not diagonal; mixing between the squark components appears on the lattice through the finite nondiagonal elements of $Z_A$. A feature of Wilson gluinos, which complicates the lattice formulation, is the appearance of an extra Yukawa coupling, $g_{Y_2}$. The ${\cal \chi} \times {\cal R}$ symmetry is broken by using Wilson discretization and thus lattice bare Green's functions are not invariant under ${\cal \chi} \times {\cal R}$ at the quantum level. In the calculation of the bare Green's functions on the lattice, we expect that mirror Yukawa terms will arise  at one-loop. The inclusion of the appropriate powers of the couplings requires the introduction of $Z_{Y_1}$ and $z_{Y_2}$, where $Z = \mathds{1} + {\cal O}(g^2)\,$ and $z = {\cal O}(g^2)$. The renormalization condition is the following: 
\begin{small}
\be
\langle   \lambda(q_1)  {\bar{\psi}} (q_2) A_+(q_3) \rangle \Big \vert^\MSbar = Z_\psi^{-1/2} Z_\la^{-1/2}  \langle   \lambda(q_1)  {\bar{\psi}} (q_2) \bigl((Z_A^{-1/2})_{++} A_+(q_3) + (Z_A^{-1/2})_{+-} A_-^\dagger(q_3)\bigr) \rangle \Big \vert^{\rm{bare}} 
\label{renormCLatt}
\ee
\end{small}
The Yukawa couplings are renormalized as follows:
\begin{small}
\be
\label{gy12}
g_{Y_1}^R \langle   \lambda(q_1)  {\bar{\psi}} (q_2) A_+(q_3) \rangle \Big \vert^\MSbar = Z_{Y_1}\, g_{Y_1}^B \langle   \lambda(q_1)  {\bar{\psi}} (q_2) A_+(q_3) \rangle \Big \vert^{\rm{bare}} + z_{Y_2}\, g_{Y_2}^B \langle   \lambda(q_1)  {\bar{\psi}} (q_2) A_+(q_3) \rangle \Big \vert^{\rm{tree}} + {\cal{O}}(g^5) 
\ee
\end{small}

Analogous equations hold for the other Green's functions which involve the Yukawa Interactions of Eq. (\ref{chiInv}) and the corresponding mirror ones in Eq. (\ref{chiNonInv}). At this point, we recall several lattice results which have been published in Ref.~\cite{Costa:2017rht}:
\begin{small}
\bea
Z_\psi^{LR,\MSbar} &=& 1 + \frac{g^2\,C_F}{16\,\pi^2} \left( -16.8025 + 3.7920 \alpha - (2+\alpha)\log\left(a^2\,\bar\mu^2\right) \right)\\
\left(Z_A^{1/2}\right)^{LR,\MSbar} &=&  \mathds{1} - \,\frac{g^2\,C_F}{16\,\pi^2}\Bigg\{\Bigg[8.1753 - 1.8960\alpha+\frac{1}{2}(1+\alpha)\log\left(a^2\,\bar\mu^2\right)\Bigg] \begin{pmatrix} 1 & 0\\ 0 & 1 \end{pmatrix} - 0.1623 \begin{pmatrix} 0 & 1\\ 1 & 0 \end{pmatrix}\Bigg\} \nonumber \\
\\
Z_{\lambda}^{LR,\MSbar} &=&  1 - \frac{g^2\,}{16\,\pi^2} \left[N_c\left(16.6444 - 3.7920 \alpha +  \alpha \log\left(a^2\,\bar\mu^2\right)\right)+ N_f\left(0.07907 + \log\left(a^2\,\bar\mu^2\right)\right) \right] \nonumber \\
\eea
\vspace{-1cm}
\bea
Z_g^{LR,\MSbar} &=&  1 + \gtilde\,\Bigg[ -9.8696 \frac{1}{N_c} + N_c \left( 12.8904  - \frac{3}{2} \log\left(a^2\,\bar{\mu}^2\right)\right)-\,N_f\left( 0.4811 - \frac{1}{2} \log(a^2\,\bar{\mu}^2)\right)\Bigg] \nonumber \\
\eea
\end{small}
At first perturbative order, ${\cal O}(g^2)\,$, Eq.~(\ref{renormCLatt}) and its counterparts involve only the difference between the one-loop $\MSbar$-renormalized and bare lattice Green's functions.  Having checked that alternative choices of the external momenta give the same results for these differences, we present it only for zero gluino momentum and with external squark field $A_+$.
\begin{small}
\begin{align}
\langle   &\lambda^{\alpha_1}(0)  {\bar{\psi}} (q_2) A_+(q_3)  \rangle^{\MSbar, {\rm{1loop}}} - \langle   \lambda^{\alpha_1}(0)  {\bar{\psi}} (q_2) A_+(q_3)  \rangle^{LR, {\rm{1loop}}} = -i\,(2\pi)^4 \delta(q_2-q_3) \frac{g_Y g^2}{16\pi^2} \frac{1}{4\sqrt2 N_c} T^{\alpha_1}\times \nonumber\\ 
\hspace{-10.75cm} &\Bigg[-3.7920 \alpha ( 1 + \gamma_5) + (-3.6920 + 5.9510 \gamma_5 +7.5840 \alpha ( 1 + \gamma_5) -  8 \gamma_5 c_{\rm hv}) N_c^2  \nonumber\\ 
&+(1 + \gamma_5) (\alpha - (3 + 2 \alpha) N_c^2 ) \log\left(a^2 \bar \mu^2 \right) 
\Bigg]
\end{align}
\end{small}
Notice that the differences between $\MSbar$-renormalized and bare lattice Green's functions with external squark filed $A_+^{\dagger}$ , $A_-$ and $A_-^{\dagger}$ contain the same decimal numbers. This is to be expected, given that these differences must lead to the same ${Z_Y}^{LR,\MSbar}$. By combining the lattice expressions with the $\MSbar$-renormalized Green's functions calculated in the continuum (see Eq.~(\ref{renormCLatt})), we find for the renormalization factors: 
\bea
{Z_{Y_1}}^{LR,\MSbar} &=& 1 + \frac{g^2}{16\,\pi^2}  \left(\frac{1.58130}{N_c} + (4.28489 - 2 c_{\rm hv})N_c + 0.520616 N_f - \frac{3}{2} C_F \log(a^2\bar\mu^2)\right) \nonumber \\
\\
{z_{Y_2}}^{LR,\MSbar} &=&\frac{g^2}{16\,\pi^2}\left(\frac{-0.08116}{N_c} + (2.49192 - 2 c_{\rm hv} )N_c \right)
\eea
We note that the above factors are gauge independent, as they should be in the $\MSbar$ scheme. Furthermore, the multiplicative renormalization $Z_{Y_1}$ is logarithmically divergent whereas the mixing coefficient $z_{Y_2}$ is finite.

	\section{Renormalization of the quartic coupling}
	\label{couplingQ}
\begin{figure}[ht!]
\centering
\includegraphics[scale=0.27]{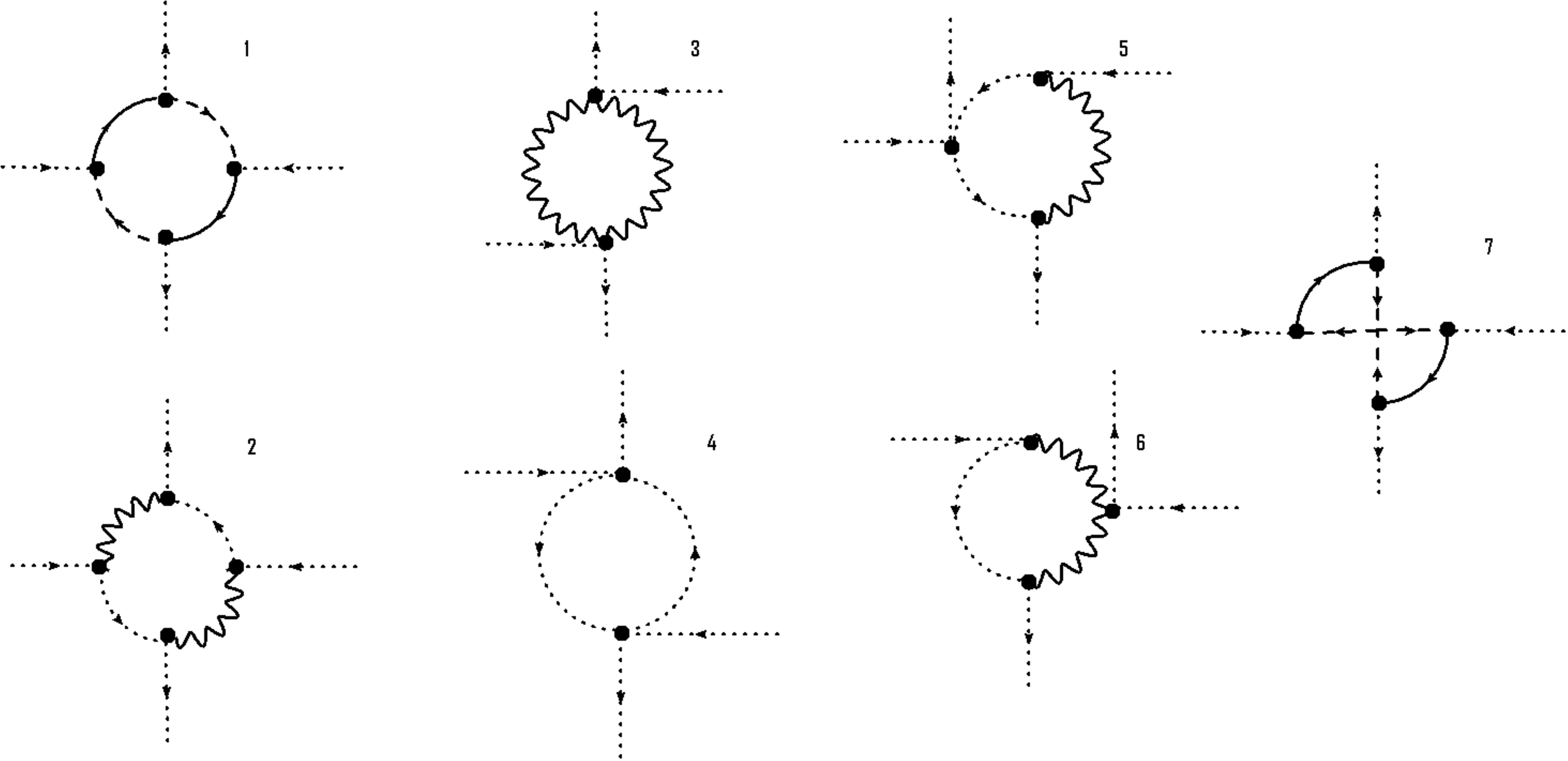}
\caption{One-loop Feynman diagrams leading to the fine tuning of $g_4$.  Notation is identical to that of Figure \ref{couplingYukawa}.
}
\label{couplingquartic}
\end{figure}

	There are seven one-loop Feynman diagrams (along with various mirror versions) that enter in the computation of the 4-point amputated Green's functions for the quartic couplings, shown in Fig. \ref{couplingquartic}. We compute, perturbatively, the relevant four-point (4-pt) Green's functions using both dimensional regularization ($DR$) and lattice regularization ($LR$).  
It is worth mentioning that the Majorana nature of gluinos manifests itself in diagram 7, in which $\lambda-\lambda$ as well as $\bar{\lambda}-\bar{\lambda}$ propagators appear.

Quartic couplings (four-squark interactions) must be appropriately fine tuned on the lattice. The $U(1)_A$ symmetry allows two squarks to lie in the fundamental representation and the other two in the antifundamental; ignoring flavor indices, there are ten possibilities for choosing the quartic couplings:
\bea
\label{externalSs}
&& (A_+^{\dagger}  A_+)( A_+^{\dagger}  A_+ ), 
\quad  (A_- A_-^{\dagger})( A_- A_-^{\dagger}), \\\nonumber &&
(A_+^{\dagger}  A_+) (A_- A_-^{\dagger}),
\quad (A_+^{\dagger}  A_-^{\dagger} )(A_+^{\dagger}  A_-^{\dagger}), 
\quad  (A_- A_+ )(A_- A_+), \quad (A_- A_+)(A_+^{\dagger}  A_-^{\dagger}),  \\\nonumber &&
(A_+^{\dagger}  A_+)( A_+^{\dagger}  A_-^{\dagger}),
\quad  (A_+^{\dagger}  A_+)( A_- A_+),  
\quad  (A_- A_-^{\dagger})( A_+^{\dagger}  A_-^{\dagger}), 
\quad  (A_- A_-^{\dagger})( A_- A_+ )
\eea
Pairs of squark fields in parenthesis denote color-singlet combinations. One must further take into account $\cal{C}$ and $\cal{P}$ to construct combinations which are invariant under these symmetries. There are five combinations as shown in Table~\ref{tb:nonsinglet2}.

\begin{table}[ht]
\begin{center}
\scalebox{0.8}{
  \begin{tabular}{c|c|c}
\hline \hline
Operators & $\cal{C}$&$\cal{P}$ \\ [0.5ex] \hline\hline
    $\la_1 [(A_+^{\dagger}  A_+)^2 + (A_- A_-^{\dagger})^2] $&$+$& $+$   \\[0.5ex]\hline
    $\la_2 [(A_+^{\dagger}  A_-^{\dagger})^2 + (A_- A_+)^2] $&$+$& $+$   \\[0.5ex]\hline
    $\la_3  (A_+^{\dagger}  A_+)  (A_- A_-^{\dagger}) $&$+$ & $+$  \\[0.5ex]\hline
    $\la_4 (A_+^{\dagger}  A_-^{\dagger}) (A_- A_+) $&$+$& $+$   \\[0.5ex]\hline
    $\la_5 (A_+^{\dagger}  A_-^{\dagger} + A_- A_+)(A_+^{\dagger}  A_+  + A_- A_-^{\dagger}) $&$+$ & $+$  \\[0.5ex]\hline
\hline
\end{tabular}}
\caption{Operators which are gauge invariant, flavor singlets and with dimensionality 4. All operators appearing in this table are eigenstates of charge conjugation, $\cal{C}$, and parity, $\cal{P}$. In the above operators, the squark fields should be explicitly identified with a flavor index. The flavor indices carried by the left fields are the same as those of right fields. The symbols $\la_i$ are the five quartic couplings.}
\label{tb:nonsinglet2}
\end{center}
\end{table}

The tree-level values of  $\la_i$ (quartic couplings) which satify SUSY are:
\be
\la_1 = \frac{1}{2}\, g^2 \, \frac{N_c -1 }{2 N_c}, \, \, \,  \la_3 =\frac{1}{2} \, g^2 \, \frac{1}{N_c}, \, \, \, \la_4 = -\frac{1}{2} \, g^2 , \, \, \, \la_2 = \la_5 =0
\ee
These couplings may receive quantum corrections, coming from the Feynman diagrams of Fig.~\ref{couplingquartic}. The calculation of these diagrams is presently underway.

	\section{Summary -- Future Plans}
	\label{summary}
	
	In this work, we study the renormalization of the Yukawa (gluino-quark-squark interactions) and the quartic (four-squark interactions) couplings that arise in $\mathcal{N} = 1$ Supersymmetric QCD. In order to extract the renormalizations and the mixing coefficients, which are related to the Yukawa coupling, we compute, perturbatively to one-loop and to the lowest order in the lattice spacing, the relevant three-point Green’s functions using both dimensional and lattice regularizations.

    In our ongoing investigation we are calculating perturbatively the relevant four-point Green’s functions so as to deduce the renormalization of the quartic coupling. It would be highly interesting to apply these fine-tunings in Monte Carlo simulations of the SQCD action. Another natural extension of our work is the perturbative study of Supersymmetric non-abelian models on the lattice using chirally invariant actions.\\
	
{\bf Acknowledgements:} 
 This work was co-funded by the European Regional Development Fund and the Republic of Cyprus through the Research and Innovation Foundation (Projects: EXCELLENCE/0918/0066 and EXCELLENCE/0421/0025). M.C. also acknowledges partial support from the Cyprus University of Technology under the ``POST-DOCTORAL" programme.

\end{document}